# Biocompatibility of Pristine Graphene Monolayers, Nanosheets and Thin Films

*Jennifer Conroy[1,3#], Navin K. Verma[2#], Ronan J. Smith[3], Ehsan Rezvani[3,4], Georg S. Duesberg[3,4], Jonathan N. Coleman[3], Yuri Volkov[1,3*]*

(1) Department of Clinical Medicine, Institute of Molecular Medicine, Trinity College Dublin, Ireland

(2) Lee Kong Chian School of Medicine, Nanyang Technological University, Singapore

(3) Centre for Research on Adaptive Nanostructures and Nanodevices (CRANN) and Advanced Materials and BioEngineering Research (AMBER) Centre, Trinity College Dublin, Ireland

(4) School of Chemistry, Trinity College Dublin, Ireland

[#] These authors contributed equally

*Corresponding Author

There is an increasing interest to develop nanoscale biocompatible graphene structures due to their desirable physicochemical properties, unlimited application opportunities and scalable production. Here we report the preparation, characterization and biocompatibility assessment of novel graphene flakes and their enabled thin films suitable for a wide range of biomedical and electronic applications. Graphene flakes were synthesized by a chemical vapour deposition method or a liquid-phase exfoliation procedure and then thin films were prepared by transferring graphene onto glass coverslips. Raman spectroscopy and transmission electron microscopy confirmed a predominantly monolayer and a high crystalline quality formation of graphene. The biocompatibility assessment of graphene thin films and graphene flakes was performed using cultured human lung epithelial cell line A549 employing a multimodal approach incorporating automated imaging, high content screening, real-time impedance sensing in combination with biochemical assays. No detectable changes in the cellular morphology or attachment of A549 cells growing on graphene thin films or cells exposed to graphene flakes (0.1 to 5 µg/mL) for 4 to 72 h was observed. Graphene treatments caused a very low level of increase in cellular production of reactive oxygen species in A549 cells, but no detectable damage to the nuclei such as changes in morphology, condensation or fragmentation was observed. In contrast, carbon black proved to be significantly more toxic than the graphene. These data open up a promising view of using graphene enabled composites for a diverse scope of safer applications.

Since monolayer graphene was first isolated and characterized a decade ago,[1] a succession of papers has described its outstanding properties in a range of areas from mechanics to optics [2]. As a result it is expected that graphene will have an impact in a wide range of applications from composites to electronics to drug delivery.[3] To bridge the gap between lab-scale demonstration and commercial development of applications, methods of producing large quantities of graphene at high production rates will be required. Such large-scale production methods can be roughly divided into two broad categories: continuous growth of monolayer graphene on substrates[4] and production of graphene nanosheets by liquid exfoliation.[5] The former method gives very high quality monolayer graphene over large areas at reasonable production rates. Importantly, the material is very well defined and consists almost solely of monolayer, sp$^2$ graphene. In contrast, graphene produced by liquid exfoliation tends to be less well-defined. The most common liquid exfoliation method involves the oxidation of graphite to give suspensions of graphene oxide (GO) nanosheets in water.[6] The resultant nanosheets have covalently bonded oxides and so display surface chemistry distinct from true graphene. As a result GO and related structures are perhaps best described as graphene-like materials. While the oxides can be removed, leading to a material known as reduced graphene oxide (rGO), structural defects always remain leading to material properties which differ greatly from true graphene.[6] Alternatively, high quality graphene can be produced from graphite in a process called liquid phase exfoliation. This results in defect- and oxide-free graphene nanosheets stabilized in solvents[7] or surfactant-water solutions[8]. While the material produced by this method is of very high quality, it tends to give few-layer rather than monolayer graphene nanosheets. This procedure has recently been scaled up to give at relatively high production rates and the process has been commercialized.[9]

This ongoing development of large-scale synthesis and the inevitable deployment of graphene-enabled products will result in the interaction of both manufacturers and consumers with graphene and so lead to the critically important question of biosafety. Being nanoscale in size and dimensions, graphene or graphene-like materials can become airborne during synthesis, processing and manufacturing. Alternatively, it is possible that graphene may detach from products during handling, manufacture, or disposal. In either case, there is a strong possibility that the interaction of graphene with humans or other animals will occur relatively frequently.

The main routes through which nanoparticles can enter the body under environmental or occupational exposure scenario are through the lungs (via inhalation), intestinal tract (via ingestion), and/or skin (transdermally) with all potentially causing adverse biological effects[10, 11, 12]. Among these exposure routes, the respiratory system is constantly exposed to the environment and is one of the most important portals of entry for nanomaterials. Once inhaled, nanosize particles can travel deep into the lung, be deposited in the alveolar region, reach detectable levels in the blood circulation after crossing the air-blood barrier, and enter specific cells/tissues, inducing toxicological effects [10, 13]. Deposition of nanomaterials in the respiratory tract can result in the formation of lung fibrosis, occurrence of multi-serous membrane fluid due to the irritation to cell membranes, development of granulomas in pleura, inflammatory infiltration and hilar lymphadenopathy[14]. In fact, a recent study using radioisotope tracing demonstrates that intratracheally instilled nanoscale graphene oxide mainly retain in the lung [15]. Thus, it becomes of utmost importance to comprehensively investigate the biocompatibility of graphene and graphene-like materials in order to address uncertainties about the health and safety and environmental implications.

A number of recent reports have described toxicity assessment of graphene and graphene-like materials. While some in vitro and in vivo studies showed no particular toxicity, others have indicated that graphene-based materials might become health hazards [16, 17, 18]. This apparent inconsistence must be addressed. To do so, it is critically important to note that graphene exists in multiple forms such as monolayer graphene, few-layer-graphene, graphene oxide, reduced graphene oxide and functionalized graphene. It should be expected that the biocompatibility will vary strongly with the graphene type. For example, a recent review by Bianco on the toxicity of graphene highlights the toxic response observed in vivo and in vitro is largely dependent on extrinsic properties such as surface chemistry.[19] Singh *et al* found that GO induced blood platelet aggregation on par with that induced by thrombin (a potent agonist of platelets) whereas rGO was significantly less effective in activating platelets[20]. In another study, Singh *et al* showed how amine-functionlised graphene did not induce any pro-thrombotic or platelet stimulating characteristics compared to GO making it safer for biomedical applications[21]. The materials described in this study (GO, rGO and Amine-GO) differ solely in their surface chemistry i.e. the presence or absence of oxides or other functional groups. These results emphasize the critical relationship between surface chemistry of graphene-like materials and toxicity. However, these surface functionalities are not intrinsic properties of pristine graphene as they are introduced in the production process. This is an important point as the toxicity such graphene-like materials may not be indicative of toxicity in intrinsic graphene. Similarly, toxicity studies on nano-scale graphenes such as nanoribbons[22] and graphene quantum dots[23] may be strongly influenced by the preponderance of edges and the associated edge chemistries in such structures.

This is an important point: the literature predominantly reports on the bio-interaction of the graphene-like materials (GO[24, 25], rGO[26], Amine-GO[21], carboxylated-GO[27], graphene nanoribbons[22] and graphene quantum dots[23]) rather than pristine graphene. The edge and surface chemistries associated with these materials are distinct from those of pristine graphene and will strongly contribute to the cellular response. The biointeraction of pristine graphene therefore remains uncertain and underexplored. We believe it is critically important to understand the intrinsic toxicity of graphene i.e. that associated with the $sp^2$ hexagonal carbon lattice. It is this that we address in this study.

Probably the material most well approximating "pure" graphene is the material exfoliated by micromechanical cleavage of graphite.[28] However, because this material will probably not be used in products we will not study its biocompatibility. A reasonable approximation to "pure" graphene which will probably be used in products is monolayer graphene produced by processes such as CVD.[4] This material can be produced as monolayers over large areas. While such material is poly-crystalline and so contains grain boundaries, it is generally free of any non-carbon functionalities and consists almost solely of $sp^2$ bonded, carbon. Alternatively, high quality few-layer graphene can be produced by liquid exfoliation as micron-sized nanosheets which are free of defects and oxides.[7, 9] Such material is produced industrially and will probably find commercial applications in areas such as printed electronics and composites.[9] Again such material consists almost solely of $sp^2$ bonded carbon and is a very good approximation of "pure" graphene.

In this work, we present the biocompatibility assessment of CVD grown monolayer graphene and surfactant-exfoliated graphene in the form of both dispersed nanosheets and thin films. Given the potential for exposure of nanomaterials to the respiratory system, we focus on this area. Therefore, we selected A549 cell line for the current studies because of its wide applicability as lung cell model system for nanotoxicity and cytocompatibility analysis[29, 30]. However, first it is necessary to characterize the graphenes used in this study in order to demonstrate their pristine state.

Graphene produced by both CVD and liquid phase exfoliation have previously been shown to be free of both oxides and basal-plane defects.[7, 9, 31] Here, we briefly demonstrate the quality of the samples used in this study. Shown in figure 1A is a transmission electron microscopy (TEM) image of a CVD graphene sheet suspended on lacey carbon grid. The diffraction pattern (shown in figure 1A inset) comprises sharp discrete six-fold symmetry spots suggesting the existence of a highly crystalline graphene film. Figure 1B shows a representative graphene flake produced by liquid phase exfoliation suspended on a holey carbon grid. Wide field TEM images can be used to generate statistics on average flake lengths for a given dispersion as shown in Fig 1C. The width of the distribution and average flake size (~0.5 μm) are typical for these kinds of dispersions.

Raman spectroscopy is a powerful tool for the characterisation of graphene based materials. There are two notable differences between the CVD graphene Raman spectrum and that of the liquid phase sample shown in Fig. 1D. The small D band observed for the CVD graphene is indicative of high crystalline quality. The larger D band seen for liquid phase samples is due to edges and is consistent with those previously reported for films with average flakes sizes between 0.5 and 1 μm[32]. The shape and relative intensity of the 2D band is also quite different. While the CVD graphene is predominantly monolayer with some bilayer regions in agreement with optical observation [33-35], the 2D band for liquid phase sample is made up of contributions from flakes with varying number of layers. This is consistent with spectra for other liquid phase exfoliated samples with mean thicknesses of 3 to 4 layers [36].

We have previously established a multimodal biohazard assessment procedure for analysing the biocompatibility of nanocomposites and reported the applicability of such a novel approach using silver nanowire enabled thin films[37]. Applying similar strategy in the current study, we assessed the potential biological effects of graphene-based thin film on human cells by performing a series of purpose-designed cellular and molecular assays using A549 cell line as a model for the inhalation exposure route. Firstly, we analysed the effect of the above described graphene thin films on the cellular morphology and cytoskeletal systems, which are important overall indicators of cell health. For this purpose, A549 cells were seeded separately on both CVD grown and exfoliated graphene thin films as well as on glass coverslips as a control. After 24 h, cells were washed, fixed and then stained for nucleus (blue), cytoskeletal proteins actin (red) and tubulin (green). Compared to the control cells on glass coverslips (Fig. 2A), no detectable changes in the gross structure of the cytoskeletal proteins actin and tubulin or the morphology of cells was observed in cells cultured on CVD grown graphene thin films (Fig. 2B) or on the exfoliated graphene thin films (Fig. 2C). The overall shapes and sizes of cells and nuclei were within the normal variation range and there were no signs of significant cellular or nuclear abnormalities (e.g., condensation or blebbing), membrane bound vesicles, shrinking of the cytoplasm or cell rupture. This is a very important result, as it shows defect- and oxide-free graphene, prepared by both CVD and liquid phase exfoliation to be biocompatible in the lung cells model system.

Considering the potential threat to workers posed by accidental occupational exposure to graphene during manufacturing, processing and transportation, it becomes essential to examine the individual graphene nanosheets from which the thin films are made. Therefore, we investigated the effect of liquid suspended graphene nanosheets on the morphology of A549 cells by cellular imaging using a confocal microscope. The A549 cells were treated with 2 μg/ml graphene for 24 h, fixed and stained for nuclei and cytoskeletal protein tubulin and actin. Similar to the morphology of the cells grown on the graphene thin films, no signs of significant cellular or nuclear abnormalities, membrane bound vesicles, or cell destruction was detected and the overall shapes and sizes of cells and nuclei were within the normal variation range (Fig. 3). Cellular accumulation of graphene was detected in treated cells (Fig. 3B).

Next we used a high throughput fluorescent microscopy technique to quantify the effect of the graphene nanosheets over a range of concentrations (1 to 5 μg/mL, selected based on their use in the fabrication of thin films) over a period of 4 to 72 h. In each of these experiments, equivalent concentrations of ultrafine carbon black were used as controls and for comparison purposes. The number of viable cells at the end of the treatment was evaluated using automated microscopy based high content screening assays and the quantitative data were converted into colour-coded heatmaps (Fig. 4A). Overall, the data clearly indicated a non-toxic response of the A549 to the graphene (Fig. 4A), whereas, a cytotoxic response was registered following ultrafine carbon black treatment at 5 µg/mL for all the time points (Fig. 4A).

Next, we evaluated the cytotoxic effect of both graphene flakes and ultrafine carbon black on A549 cells by monitoring them in real-time following treatment for up to 96 h using a live cell-based electrical impedance sensing technique. Cells were allowed to grow on the bottom of the wells, each containing electrodes, to confluence over a period of 20 h, and then either 5 µg/mL graphene or carbon black were added. Changes in resistance of the individual electrodes, depending on various morphological parameters of cells affected by these treatments, were quantified automatically and plotted as an arbitrary unit Cell Index (Fig. 4B). No detectable toxicity was observed due to graphene treatment, whereas a moderate but significant level of cytotoxicity was registered following carbon black exposure in comparison to the untreated controls.

To further determine the effect of graphene on A549 cell viability, we performed an alternative cell metabolism-based assay quantifying the conversion of MTS (tetrazolium salts) to formazan. No significant decrease in cell viability was detected following graphene exposure even at the highest concentration of 5 µg/mL over 24 h (Fig. 4C). However, cell death of 24.1% was observed when A549 cells were treated with 5 µg/ml carbon black for 24 h (Fig. 4C).

It is well documented that toxic agents often raise the intracellular level of reactive oxygen species (ROS), which is one of the main characteristics of cellular oxidative stress and is an important indicator of the cell health[12, 38]. In the present study we analyzed if graphene treatments can induce intracellular ROS levels in A549 cells by HCA using a fluorescent indicator DCFH-DA. A very low level of increase in ROS generation in A549 cells was detected following treatments with various concentrations of graphene ranging from 0.1 µg/mL to 5 µg/mL (Fig. 5). In contrast, cells exposed to 100 µM $H_2O_2$ as a positive control showed significantly elevated ROS levels (Fig. 5A).

We further explored the effect of graphene on the nuclear phenotype. No significant changes in the nuclear morphology or intensity of nuclear staining were observed in cells treated with graphene as compared to untreated control (Fig. 5B). In contrast, cells treated with staurosporine as a positive control displayed condensed nuclei with very high levels of nuclear staining intensity, a classical morphological feature of apoptosis (i.e. reduction in size, chromatin condensation, and DNA fragmentation). Moreover, no TUNEL-positive A549 cells were detected following graphene treatment (Fig. 5C), while significant number of staurosporine treated cells showed TUNEL staining. No signs of histone phosphorylation, which is one of the most important signatures of nuclear damage, was found following graphene exposure in A549 cells (Fig. 5D).

In this study, we systematically investigated two different scenarios of cell exposure to graphene - in one case A549 cells were incubated on transparent graphene thin films while in the second scenario cells were exposed to various concentrations of graphene flakes (ranging from 0.1 to 5 µg/mL) in a loose unbound state. We believe that the biological/toxicological effects of nanomaterials should be evaluated using multimodal assays combining both biochemical and advanced microscopic techniques. Therefore, special considerations have been given to the choice of techniques and the design of experiments to analyse the potential effect(s) of graphene on biological systems. Here we implemented automated image acquisition and multi-parametric analysis in combination with real-time impedance sensing and a standard MTS-based cell viability assay. Both graphene flakes as well as their enabled thin films were found to be biocompatible under the current experimental conditions.

Further, we employed a cell-based high content screening technique that operates on the principle of fully automated fluorescence microscopy and is considered as a useful tool in modern drug discovery and in nanotoxicity evaluation [30, 37, 39-41]. This technology is particularly well-suited to studying cytotoxicity of nanomaterials, because it allows for multiplexing of key reporter parameters such as cell morphology and viability to be analyzed for a large number of samples varied experimental conditions in a high throughput manner. Impedance sensing method permits environmental control and very closely imitates in vivo physiological conditions [37]. MTS-based colorimetric measurement of mitochondrial metabolic rates reflects cell viability and is widely accepted as a standard way to examine the effect of toxic agents on cellular systems. The combined use of multi-modal biohazard assessment platform implemented in the current study allows identifying a far wider scope of toxicological phenomena that may hold true at the nano-bio interface and offers mechanistic interpretations that can be exploited to better interrogate cellular responses to nanoparticles and nanomaterials.

Multiple studies have reported that nanomaterials of various origins and types can cause non-specific oxidative damage resulting in their cytotoxicity[15, 42, 43] and some nanoparticles are capable of acting as pro-oxidants due to the chemical properties of their surface. Reactive oxygen species (ROS) were reported to cause nuclear damage, which contributes to apoptotic cell death in a variety of cell types [43, 44]. In the current experimental conditions, a very low level of increase in ROS was recorded, but no signs of nuclear damage in terms of nuclear morphology, condensation or fragmentation were detected. It may be possible that graphene absorbs some of the nutrients or serum proteins in culture medium [45] and then their depletion induces the observed low level of oxidative stress response to A549 cells. Such mechanism has previously been reported in the study on carbon nanotubes [18, 46], where the depletion of nutrients by their absorption onto the nanotubes caused severe toxicity to HepG2 cells [18]. It may also be possible that the cells can withstand such as low or negligible levels of stress caused by ROS and this explains no detectable damage to the nuclei. Of note, we have previously observed that nanomaterials such as single-walled carbon nanotubes can be also biodegraded intracellularly[47].

Unlike a number of previous reports by various research groups pertaining to the toxicity of graphene derivatives on different model systems [16, 25, 48], we found that A549 cells grew very well on both CVD-grown graphene monolayers and thin films of graphene nanosheets. This apparent inconsistency in the cytocompatability of graphene almost certainly comes from the type of graphene used. All the previous studies which reported toxicity used chemically modified graphene-like materials such as graphene oxides containing many oxygen atoms in the forms of carboxyl groups, epoxy groups and hydroxyl groups [49]. Here, when using graphene which is free of oxides or other functional groups we see very good biocompatibility. This strongly implies graphene to be intrinsically non-toxic with toxicity potential only appearing after chemical treatment. We believe that surface chemistry is the biggest factor controlling the biocompatability and note that it controls a number of other important factors such as surface energy and hydrophobicity. However, a number of other influences might contribute to the mechanistic (aggregation, cellular processes, biodistribution, and degradation kinetics) and toxicological outcomes of the graphene-based materials with the obvious examples being layer number, lateral size, stiffness and impurity/defect content[19, 50]

In conclusion, we present here the results of our continuous efforts in the development of a novel biocompatible graphene and graphene thin films. The cytotoxicity potential of our nanocomposite preparations was investigated using a multimodal assessment approach. Overall, our data suggests that the developed pristine graphene flakes and thin films are biocompatible at the explored concentration range and can further be utilized for a wide range of safer biotechnological, medical and electronics applications.

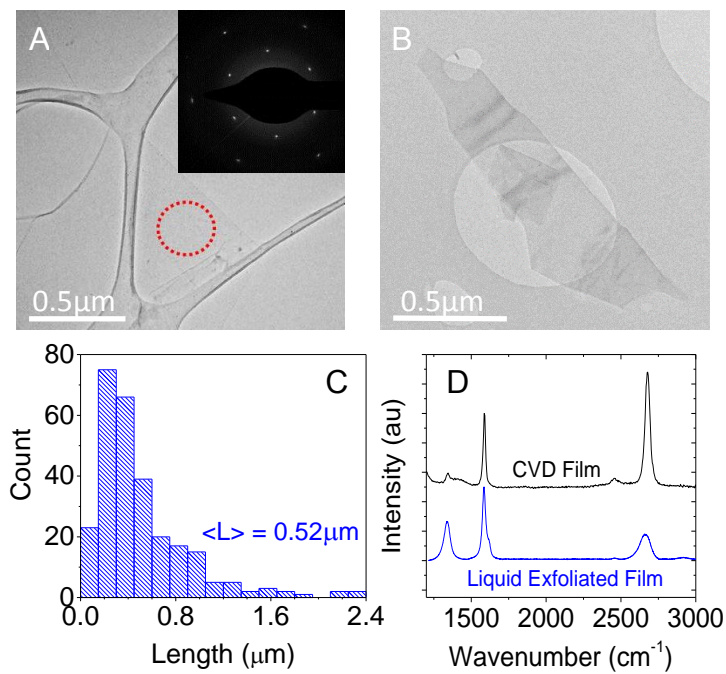

**Figure 1 Representative Raman spectra and TEM images of CVD/Liquid Exfoliated graphene. A)** TEM image of a CVD graphene sheet suspended on lacey carbon grid indicating presence of large-area continuous graphene. The inset displays the diffraction pattern which consists of sharp discrete six-fold symmetric spots suggesting the existence of a highly crystalline graphene with low density of defects. The red dashed circle indicates where diffraction pattern was taken. **B)** Representative TEM image of graphene sheet produced in by liquid phase exfoliation suspended on a holey carbon grid **C).** Size distribution of graphene flakes dispersed in sodium cholate measured using TEM. Average flake length was ~500nm. **D)** Average Raman Spectra for CVD graphene sheet and thin film cast from liquid phase graphene flakes. Spectra are normalised to G peak intensity.

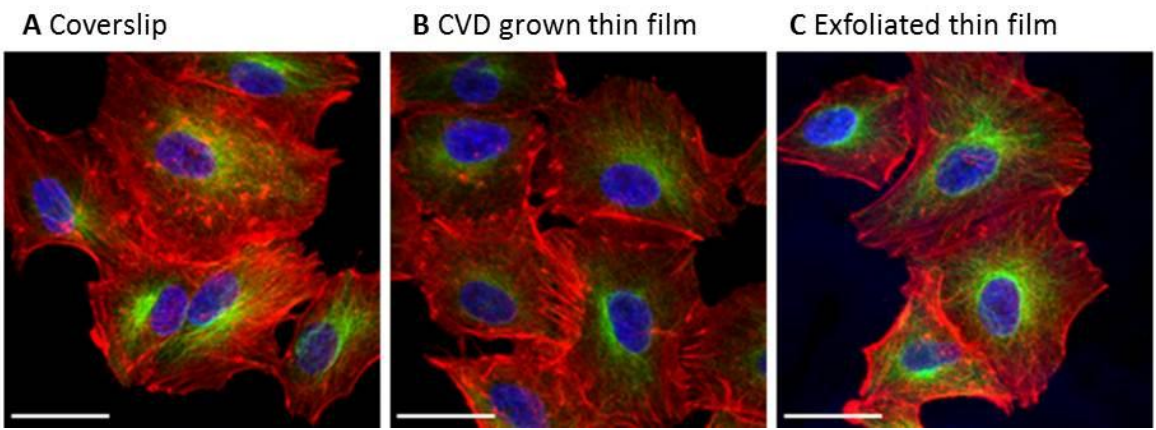

**Figure 2 Confocal microscopy images of A549 cells growing on graphene thin films. A549 cells were incubated on a glass coverslip (A), CVD-grown graphene thin film (B) or exfoliated graphene thin film (C) for 24 h. Cells were fixed, counterstained for cytoskeletal proteins α-Tubulin (green), Actin (red), and nuclei (blue), and imaged by a confocal microscope (63x oil). Images display normal cellular and nuclear morphology and cytoskeletal systems. Sale bar 20 μm.**

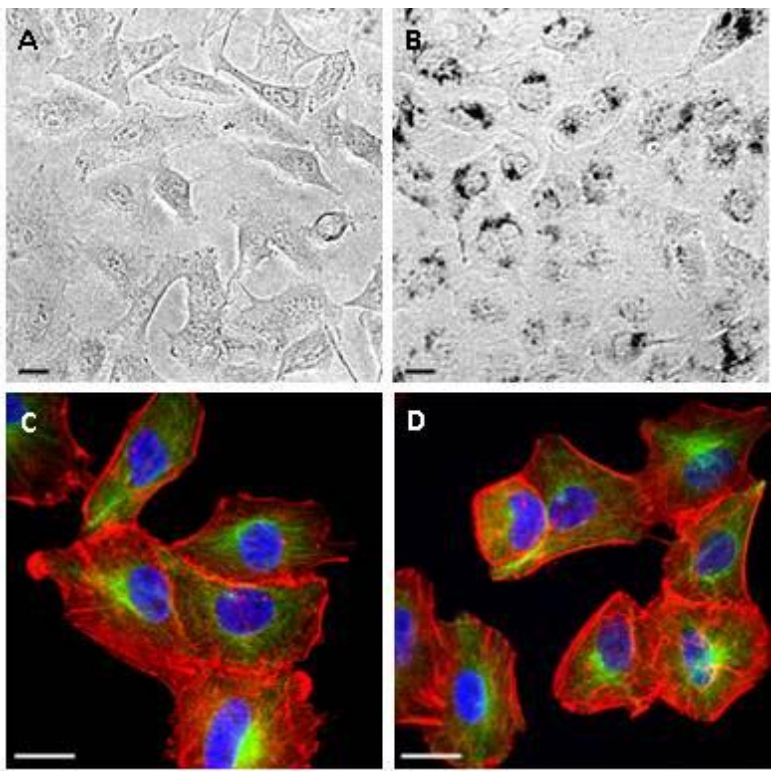

**Figure 3 Effect of graphene flakes on cellular morphology or cytoskeletal systems of A549 cells.** A549 cells growing on a glass coverslips were untreated (A, C) or treated with graphene flakes (2 µg/mL) for 24 h (B, D) and fixed. Cells were counterstained for cytoskeletal proteins α-Tubulin (green), Actin (red), and nuclei (blue), and imaged by a confocal microscope (63x oil). Graphene treated cells display normal cellular attachment and morphology. Cellular accumulation of graphene can be seen in the brightfield image of treated cells (B) while fluorescent images indicate normal nuclear morphology and cytoskeletal systems. Sale bar 20 µm.

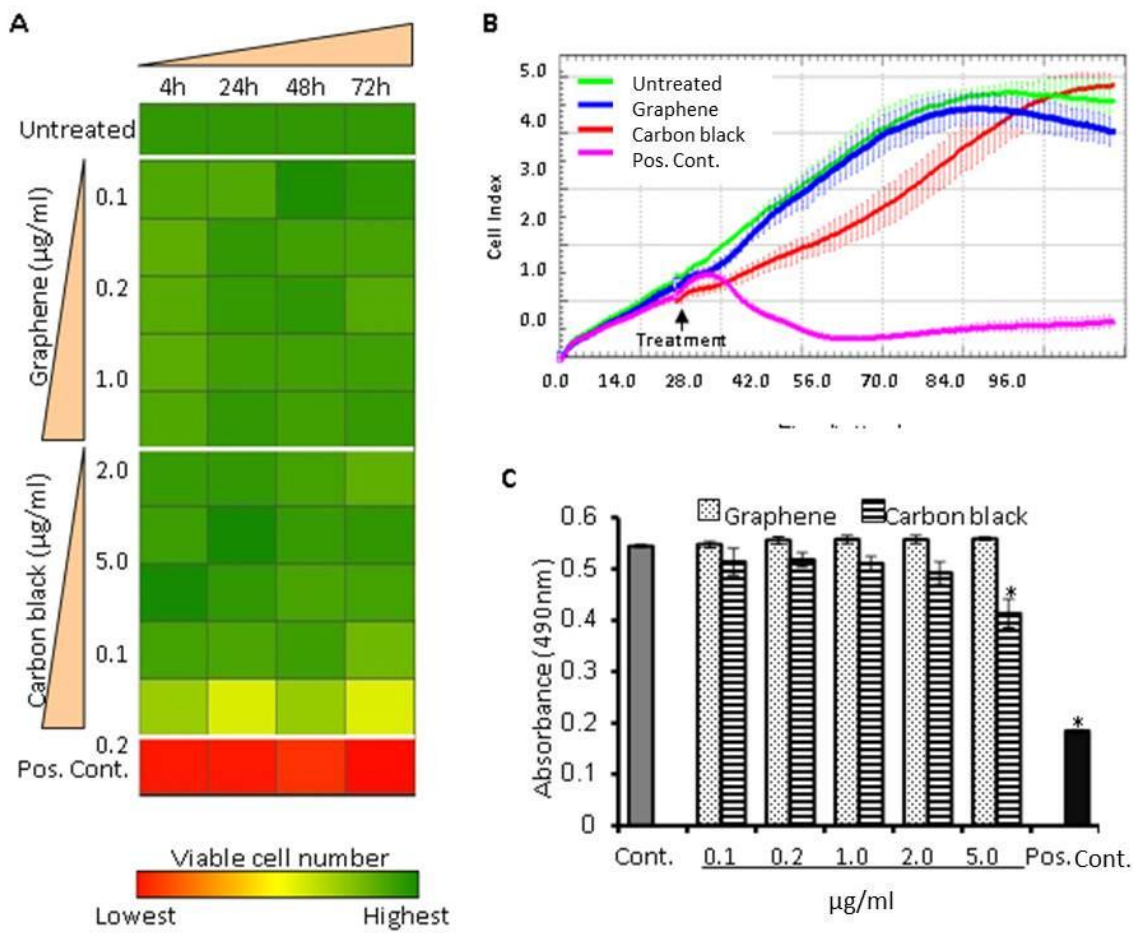

**Figure 4 Comparative cytocompatibility analysis of graphene vs carbon black in A549 cells. (A) A549 cells ($5 \times 10^3$ cells/well) growing on 96-well tissue culture plates were incubated with the indicated concentration of graphene flakes or ultrafine carbon black (0.1, 0.5, 1, 2 or 5 μg/mL) for 4 to 72 h. Cells were also treated with a cytoskeletal disrupting drug nocodazole (5 μg/mL) as a positive toxicity control (Pos. Cont.). Cytocompatibility analysis was performed using an automated microscope IN Cell Analyzer-1000 equipped with Investigator software by quantifying cell adherence to the plates. A heatmap based on the normalized cell numbers of three independent experiments in triplicates from five randomly selected fields per well containing at least 300 cells was generated and presented. (B) A549 cells (5 x 103 cells/well) growing on 96-well E-plates for 20 h were treated with graphene (2 μg/mL), carbon black (2 μg/mL) or nocodazole (5 μg/mL, Pos. Cont.) in triplicate and incubated for additional 76 h. Cell growth curves (Cell Index ± SEM) were automatically recorded by dynamic monitoring of cell adhesion and spreading process using xCELLigance electrical impedance sensing platform. (C) Cells were cultured in 96-well tissue culture plates, treated with graphene flakes (0.1, 0.5, 1, 2 or 5 μg/mL), or nocodazole (5 μg/mL, as a positive toxicity control) for 24 h and cell viability was evaluated by MTS-based assay. Data represents three independent experiments performed in triplicate samples. *$p < 0.05$ compared to untreated control.**

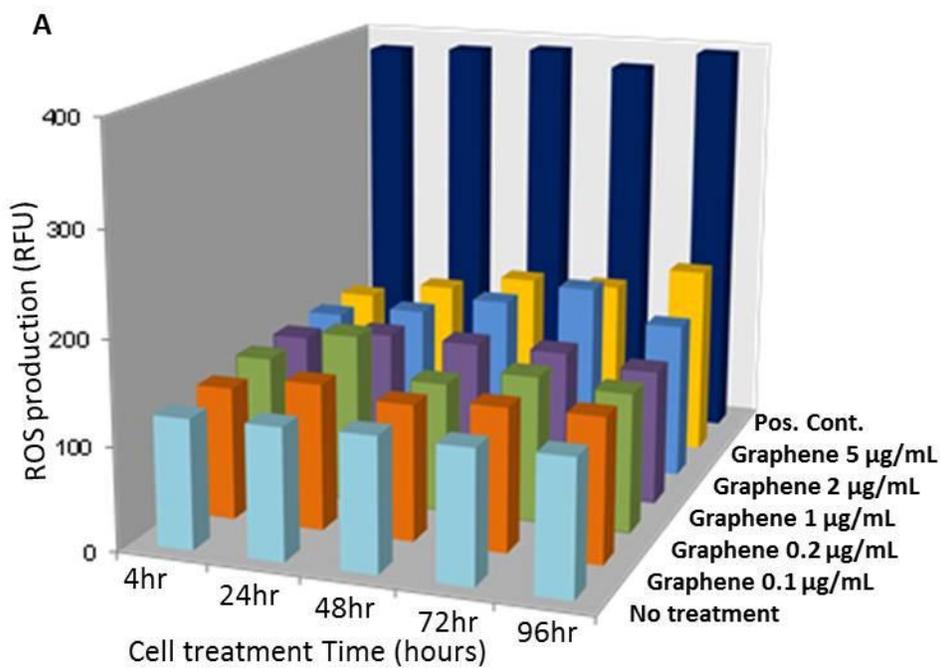
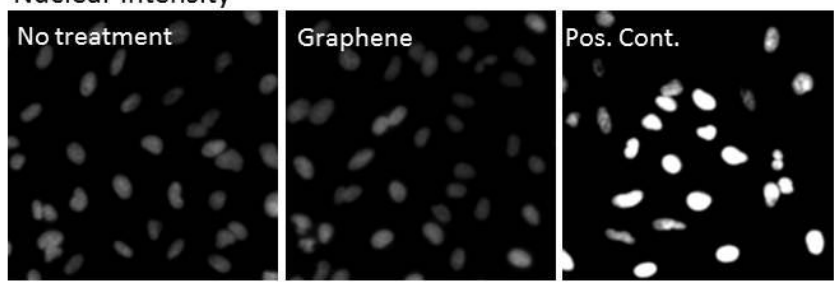
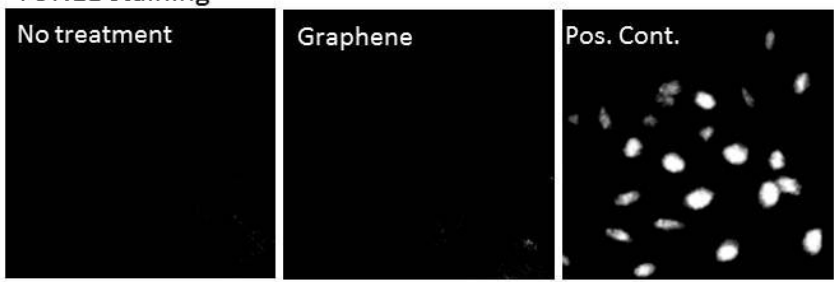
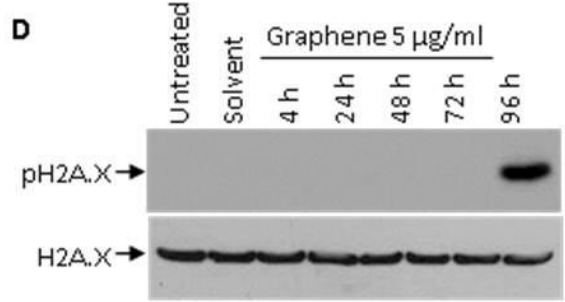

**Figure 5 Effect of graphene on cellular ROS generation, nuclear morphology, fragmentation and phosphorylation of histone.** A549 cells seeded in 96-well plates (5 x $10^3$ cells/well) in triplicates were allowed to adhere by incubating for 24 h. (A) Cells were then treated with various concentrations of graphene flakes (0.1, 0.2, 1, 2 or 5 µg/mL) for a range of time points (4 h, 24 h, 48h, 72 h, or 96 h) or 100 µM $H_2O_2$ for 1 h (as a positive control, Pos. Cont.). At the end of the treatment periods cells were probed with DCFDA and plates were scanned using IN Cell Analyzer 1000 automated microscope. ROS generation was automatically quantified using IN Cell Investigator software and presented. (B) Cells were then treated with graphene flakes (5 µg/mL) or staurosporine (100 ng/mL as a positive control, Pos. Cont.) for 24 h. At the end of the treatment period adherent cells were fixed, stained with Hoechst and nuclear images were acquired using IN Cell Analyzer 1000 automated microscope. (C) Cells were subjected for TUNEL staining and imaged using IN Cell Analyzer 1000. (D) Cells cultured in 6-well plates (5x $10^5$ cells/well) were treated with 5 µg/mL graphene for 4 h, 24 h, 48 h, 72 h or 96 h, or 100 ng/ml staurosporine for 24 h and lysed. Cellular lysates were resolved by SDS-PAGE and Western immunobloted for phospho-histone or histone. Results are representative of three independent experiments.

## Materials and Method

*Graphene preparation*

Graphene was prepared using two methods namely chemical vapour deposition (CVD) and liquid-phase exfoliation. CVD-grown graphene films were prepared on 18 µm thick copper foils (99.8% Gould Electronics GmbH) using methane as carbon feedstock which has been reported in detail elsewhere [51]. In brief, samples were heated in a tube furnace (Carbolite) to 1035 °C under $H_2$ flow (50 sccm). The Cu foil was annealed for an hour at this temperature to remove any oxides and increase the Cu grain size. Then a mixture of $CH_4$ (10 sccm) and $H_2$ (2.5 sccm) was introduced for 20 min to grow graphene. The samples were then cooled to room temperature under hydrogen flow. Subsequently, the graphene film was transferred onto standard lab glass cover slips using the established polymer-assisted transfer technique[35]. Extra care was taken and extensive rinsing steps were implemented to minimise contamination. A stock dispersion of liquid phase exfoliated graphene was prepared by adding graphite powder (Aldrich) at initial graphite concentration 20 mgml$^{-1}$ to sodium cholate solution 0.3 mgml$^{-1}$ as described previously. Ultra sonication was carried out using a high powered sonic tip (Sonics VX-750 ultrasonic processor) for approximately 4 h. The dispersion was left to sit overnight and then centrifuged (Hettich Mickro 22R) at 1500 rpm for 90 mins to remove large aggregates. Dispersed concentration was calculated using optical absorption measurements (taking $\alpha_{660\ nm} = 6600$) performed on a Varian Cary 6000i UV–Vis–IR absorption spectrometer.

*Raman spectroscopy*

Raman spectroscopy was used to assess the quality and uniformity of the transferred graphene films on glass cover slips substrate. Raman measurements were performed using a Witec Alpha 300R with a laser excitation wavelength of 532 nm. A number of areas across the sample were mapped out to ensure the data are representative of the whole sample.

Raman spectroscopy is a powerful tool for the characterisation of graphene based materials. The most prominent features are the G band at ~1580 cm-1 and the 2D band at ~2660 cm-1. When disordered regions of graphitic samples are probed, e.g. edges, vacancies, defects etc., a disorder induced D band manifests at ~1350 cm-1 [52]. The 2D peak is particularly useful when characterising graphene as its intensity, position and line shape change depending on the number of layers present [34]. The 2D peak for monolayer graphene is typically observed at ~2660 cm-1 and can be well fitted with a single Lorentzian peak with a FWHM < 30 cm-1. Additionally, the 2D to G peak intensity ratio (I2D:IG) is typically ~3 for pristine monolayer graphene samples. In the case of bilayer graphene the 2D peak is blue shifted and can be fitted with four distinct contributions, broadening its FWHM and suppressing its I2D:IG (typically <1). As additional layers are added, the Raman spectrum changes towards that of bulk graphite. [33, 53]. The average spectrum shown in Fig. 1B shows a symmetric 2D peak with FWHM of ~36 cm-1 and I2D: IG of 1.8 (Fig. 1D)suggesting that the CVD graphene is predominantly monolayer with some bilayer regions, which is in agreement with optical observation [33-35]. A small D band is observed indicating that the CVD graphene is of high crystalline quality. There are two notable differences between the CVD Graphene Raman spectrum and that of the liquid phase sample. The appearance of a large D band would seem to suggest increased disorder in the film compared with the CVD sample. This can be easily understood by realising that the film is made up of a disordered array of flakes rather than a continuous graphene lattice. As such the contribution from edge defects increases the ID:IG ratio. The D band seen here is consistent with those previously reported for films with average flakes sizes between 0.5 and 1 μm. The shape and relative intensity of the 2D band is also quite different. This is due to contributions from flakes with varying number of layers and is consistent with spectra for other liquid phase exfoliated samples.

*Transmission electron microscopy (TEM)*

In order to further evaluate the microstructure and crystalline quality of the films, CVD-grown graphene on copper foil was transferred directly onto transmission electron microscopy (TEM) grids using a polymer-free technique [54]. TEM studies were performed in an FEI Titan 80-300 kV S/TEM. Liquid Phase samples for TEM analysis were prepared by drop-casting the dispersion onto holey carbon grids (400 mesh). Bright-field TEM images were taken with a JEOL 2100, operated at 200 kV.

The TEM image of a CVD graphene sheet suspended on lacey carbon grid indicates the presence of large-area continuous graphene (Fig. 1A). The diffraction pattern comprises sharp discrete six-fold symmetry spots suggesting the existence of a highly crystalline graphene film. Figure 1C shows a representative graphene flake produced by liquid phase exfoliation suspended on a holey carbon grid. Although centrifugation removes large aggregates the graphene is still polydispersed and hence each dispersion contains a distribution of flake sizes and thickness. Wide field TEM images can be used to generate statistics on average flake lengths for a given dispersion as shown in Fig 1F. The width of the distribution and average flake size (~0.5 μm) are typical for these kinds of dispersions.

*Zeta Potential Measurements*

Zeta potential measurements were carried out on a Malvern Zetasizer Nano system with irradiation from a 633 nm He–Ne laser. The samples were injected into folded capillary cells, and the electrophoretic mobility (μ) was measured using a combination of the electrophoresis and laser Doppler velocimetry techniques. The electrophoretic mobility relates the drift velocity of a colloid (v) to the applied electric field (E) : $v = \mu E$. All measurements were carried out at 20°C and at the natural pH of the surfactant solution

Surfactant-coated nanoparticles are usually stabilized by electrostatic repulsion as characterized by the zeta potential. At neutral pH, the zeta potential of the sodium cholate coated graphene was measured to be stable over weeks at −50 mV, well above the minimum value for colloidal stability. The spectrum is broad and asymmetric due to the range of flake sizes as well as possible contributions from free surfactant.

*Cell culture and treatments*

The human alveolar epithelial cell line A549 (European Collection of Cell Cultures Salisbury, England), which is a well-characterised and widely used model system for comprehending pulmonary nanotoxicity[29], was cultured as described previously[30]. Briefly, cells were cultured in Gibco® Ham's F12 medium supplemented with 10% (v/v) fetal bovine serum, 50 U/ml penicillin and 50 μg/ml streptomycin in a humidified incubator at 37°C and 5% $CO_2$. All the cell culture reagents were obtained from Life Technologies Corporation (Bio-Sciences, Dublin, Ireland). For experimentation, cells were seeded into Nunc® 6-well or 96-well tissue-culture plates (Fisher Scientific Ireland Ltd., Dublin, Ireland) at a density of $5 \times 10^5$ or $5 \times 10^3$ cells/well and allowed to grow overnight prior to treatment. For studies involving thin films, cells were seeded onto the films placed at the bottom of the Nunc® 96-well plates. Graphene stock suspensions were diluted in cell culture medium followed by vigorous vortexing and immediately applied to the cells.

*Confocal microscopy*

For confocal microscopy, A549 cells were cultured on graphene thin films or glass cover-slips for 24 h, or treated with graphene flakes as appropriate. At the end of the treatment, cells were fixed in 3% paraformaldehyde and then washed with phosphate buffered saline (PBS). Cells were stained with fluorescently labelled with FITC conjugated α-tubulin (Sigma-Aldrich Ireland Ltd., Wicklow, Ireland) and Alexa Fluor 569 phalloidin (Molecular Probes®, Life Technologies Corporation, Bio-Sciences, Dublin, Ireland) to visualize the cytoskeleton systems and Hoechst (Sigma-Aldrich Ireland Ltd., Wicklow, Ireland) to visualize the nuclei. Coverslips or thin films were mounted on glass slides using mounting media (Dako). Confocal imaging was carried out by a laser scanning Zeiss LSM510-Meta microscope (Carl Zeiss Microimaging Inc., NY, USA) using a ×63 oil immersion objective lens. Excitation wavelengths used were 405 nm, 488 nm and 561 nm and emission filters were BP 420-480 nm, BP 505-530 nm and 572-754 nm respectively. At least 20 different microscopic fields were analyzed for each sample.

*High content screening (HCS) and analysis*

HCS protocols have been optimized and established in our laboratory as described previously [30, 37, 41, 55]. Briefly, A549 cells were seeded in Nunc® 96-well tissue culture plates and after 24 h treated with various concentrations of graphene flakes (ranging from 0.1 to 5 µg/mL, selected based on their use in the fabrication of thin films) or equivalent concentrations of carbon black (for comparative toxicity analysis) for 4 to 72h h. Cells were washed with PBS and fixed in 3% paraformaldehyde. Cells were stained with fluorescently labelled with FITC conjugated α-tubulin (Sigma-Aldrich Ireland Ltd., Wicklow, Ireland) and Alexa Fluor 569 phalloidin (Molecular Probes®, Life Technologies Corporation, Bio-Sciences, Dublin, Ireland) to visualize the cytoskeleton systems and Hoechst (Sigma-Aldrich Ireland Ltd., Wicklow, Ireland) to visualize the nuclei. The 96-well plates were scanned (9 randomly selected fields/well) by an automated microscope IN Cell Analyzer 1000 (GE Healthcare, UK) and images were acquired using ×10 or ×20 objectives. Quantitative estimations of the acquired images were performed with IN Cell Investigator software using multi-parameter cytotoxicity bio-application module (GE Healthcare, UK). The normalized data were converted into heatmaps using Spotfire (TIBCO Software Inc., Somerville, MA).

*Real-time impedance sensing*

To monitor the effects of graphene on A549 cell viability and proliferation in real-time, we performed electrical impedance based cytotoxicity assay using xCELLigance system as per manufacturer's instructions (Acea Biosciences Inc.) and described previously[41, 55]. Briefly, cells were seeded at a density of $5 \times 10^3$ cells/well into 100 µl of media in the E-Plate® (cross interdigitated micro-electrodes integrated on the bottom of 96-well tissue culture plates by micro-electronic sensor technology) and allowed to attach onto the electrode surface over time. The electrical impedance was recorded every 15 minutes. At the 20 h time point, when cells adhered to the well properly, they were treated with graphene or carbon black in triplicates and monitored for additional 76 h. The cell impedance (which depends on cell number, degree of adhesion, spreading and proliferation of the cells), expressed in arbitrary units called the 'Cell

Index', were automatically calculated on the xCELLigence system and converted into growth curves.

*Reactive oxygen species (ROS) measurement*

The intracellular generation of ROS was measured using a fluorescent probe 6-carboxy-2,7′-diclorodihydrofluorescein diacetate, di(acetoxy ester) (DCFH-DA) (Molecular Probes, Life Technologies Corporation, Bio-Sciences, Dublin, Ireland) as per manufacturers' instructions and with minor modifications. Briefly, cells pre-treated with increasing concentrations of graphene (0.1, 0.2, 1, 2, or 5 μg/ml) for a range of time points (4, 24, 48, 72, or 96 h) were incubated with 5 μM of DCFH-DA for 1 h. Cells untreated or treated with hydrogen peroxide ($H_2O_2$, 100 μM) for 1 h were also stained with DCFH-DA and taken as negative or positive control. The cells were washed with PBS, scanned using IN Cell Analyzer 1000 automated microscope and analyzed by IN Cell Investigator software (GE Healthcare, UK).

*TUNEL assay*

To analyze cell-death by apoptosis, we performed Terminal deoxynucleotidyl transferase dUTP nick end labeling (TUNEL) assay according to manufacturer's instructions (ApoAlert® DNA Fragmentation Assay Kit, Clontech Laboratories, Inc., Mountain View, CA) and as described[40]. Fluorescent images were acquired and analyzed using IN Cell Analyzer 1000 automated microscope and IN Cell Image Investigator software (GE Healthcare, UK).

*MTS-based cell viability assay*

Cell viability was determined using CellTier 96® AQueous One solution cell proliferation assay kit according to the manufacturer's instruction (Promega Corporation, Madison, WI) and described previously [55]. This assay evaluates the mitochondrial functioning by measuring ability of viable cells to reduce MTS into blue formazon product. Briefly, at the end of the treatment period, the 96-well plate containing 100 μl of cell culture was incubated with 20 μl of MTS tetrazolium solution (provided by manufacturer) for 4 h at 37ºC. Subsequently, the absorbance was measured at 490 nm using a microplate reader (Tecan, Mannedorf, Switzerland) and then relative cell viability was calculated. Each treatment was performed in triplicate.

*Western blot analysis for detection of pH2A.X*

The A549 cells untreated or treated with graphene flakes for various time points were lysed in lysis buffer as described previously[56]. The protein content of the cell lysates was determined by Bradford assay. Cellular lysates were resolved by sodium dodecyl sulphate polyacrylamide gel electrophoresis (SDS-PAGE). After blocking with 5% non-fat milk, the membranes were incubated overnight at 4°C with a primary antibody against phospho-H2AX (pH2A.X), or H2AX, then with a secondary antibody conjugated with HRP for 1 h. All these antibodies were from Cell Signalling Technology and used at 1:1000 dilutions. The immunoreactive bands were visualized using the chemiluminescence detection system (Cell signalling Technology, Danvers, MA) and subsequent exposure to Kodak light sensitive film (Cedex, France).

*Statistical analysis*

The data are expressed as mean ± standard error of mean (SEM). For comparison of two groups, *p*-values were calculated by two-tailed unpaired student's t-test. In all cases *p*-values < 0.05 was considered to be statistically significant.

**Corresponding Author**

# email: Yuri Volkov, volkovy@tcd.ie

**Author Contributions**

*These authors contributed equally

**Acknowledgements**

The authors would like to acknowledge SFI funded CRANN-HP collaboration (Contracts No. 08/CE/I1432S) for financial support. This work was also supported in part by EC FP7 NMP Project NAMDIATREAM (ref. 246479), the Higher Educational Authority of Ireland (HEA) and AMBER.